\documentclass[preprintnumbers,twocolumn,prl,aps,superscriptaddress,footinbib,amsfonts,showpacs,floatfix]{revtex4-1}

\usepackage[T1]{fontenc}
\usepackage{blindtext}
\usepackage{amsmath}
\usepackage{amssymb}
\usepackage{graphics,graphicx}
\usepackage{dcolumn}
\usepackage{bm}
\usepackage{epsfig}
\usepackage[normalem]{ulem}
\usepackage[usenames]{color}
\usepackage{mathbbol}
\usepackage{epstopdf}
\usepackage{simplewick}
\usepackage[utf8]{inputenc} 
\usepackage{slashed}
\usepackage{soul}
\usepackage[dvipsnames]{xcolor}
\usepackage[export]{adjustbox}
\usepackage{mathtools}
\usepackage[colorlinks = true, citecolor = blue, linkcolor = blue, urlcolor = blue]{hyperref}

\allowdisplaybreaks

\begin{document}

\preprint{}

\title{\bf Dihadron fragmentation framework for near-side energy-energy correlators}

\newcommand*{\UCLA}{Department of Physics and Astronomy, University of California, Los Angeles, CA 90095, USA}\affiliation{\UCLA}
\newcommand*{\Bhaumik}{Mani L. Bhaumik Institute for Theoretical Physics, University of California, Los Angeles, CA 90095, USA}\affiliation{\Bhaumik}
\newcommand*{\SB}{Center for Frontiers in Nuclear Science, Stony Brook University, Stony Brook, NY 11794, USA}\affiliation{\SB}
\newcommand*{\TU}{Department of Physics, SERC, Temple University, Philadelphia, Pennsylvania 19122, USA}\affiliation{\TU}
\newcommand*{\LVC}{Department of Physics, Lebanon Valley College, Annville, Pennsylvania 17003, USA}\affiliation{\LVC}

\author{Zhong-Bo Kang}\email{zkang@physics.ucla.edu}\affiliation{\UCLA}\affiliation{\Bhaumik}\affiliation{\SB}
\author{Andreas Metz}\email{metza@temple.edu}\affiliation{\TU}
\author{Daniel Pitonyak}\email{pitonyak@lvc.edu}\affiliation{\LVC}
\author{Congyue Zhang}\email{maxzhang2002@g.ucla.edu}\affiliation{\UCLA}\affiliation{\Bhaumik}

\begin{abstract}
\noindent We establish an approach to analyze the free hadron and transition (nonperturbative) regions of near-side energy-energy correlators (EECs) based on dihadron fragmentation functions~(DiFFs).  We introduce a (nonperturbative) function we call the ``EEC-DiFF'' and explicitly show that expanding it for large relative transverse momentum between the two hadrons gives the $\mathcal{O}(\alpha_s)$ expression for the ``EEC jet'' function used in the quark/gluon (perturbative) region.  This connection indicates that a formal theoretical matching will be able to bridge the free hadron, transition, and quark/gluon regions and allow all of them to be analyzed simultaneously.  We further derive a result valid for near-side EECs in the free hadron and transition regions of $e^+e^-$ annihilation in terms of the EEC-DiFF.  Using a simple model for the function, we perform the first fit within the dihadron framework to experimental data in this regime.  
We find reasonable agreement with the measurements and reproduce the salient features of near-side EECs in the free hadron and transition regions.
\end{abstract}

\maketitle 

{\it Introduction and Motivation---} The theory of the strong interaction, quantum chromodynamics (QCD), is characterized by two defining features:~asymptotic freedom~\cite{Gross:1973id,Politzer:1973fx} and confinement~\cite{Wilson:1974sk}.  The former allows for collisions with a hard energy scale to be described by quarks and gluons (partons), whereas the latter conveys the fact that after such collisions, partons will hadronize into colorless bound states.  Measurements of energy-energy correlators (EECs) as a function of the opening angle $\chi$ between the two hadrons, an observable first proposed in the 1970s~\cite{Basham:1977iq,Basham:1978bw,Basham:1978zq,Basham:1979gh}, display clear signatures for these quark/gluon (perturbative) and free hadron (nonperturbative) regions with a continuous transition between them~\cite{CELLO:1982rca,Fernandez:1984db,JADE:1984taa,PLUTO:1985yzc,Wood:1987uf,TASSO:1987mcs,TOPAZ:1989yod,ALEPH:1990vew,OPAL:1990reb,OPAL:1991uui,L3:1991qlf,L3:1992btq,SLD:1994idb,SLD:1994yoe,Bossi:2025xsi,CMS:2024mlf,ALICE:2024dfl}. (For an extensive review on energy correlators, see Ref.~\cite{Moult:2025nhu}.)  

Therefore, a theoretical framework that simultaneously captures the different characteristics of EECs in these regions can offer great insight into our understanding of QCD.  
In particular, two angular regimes of EECs have been of interest:~near side (or collinear) $\chi\approx0$~\cite{Dixon:2019uzg,Chen:2023zzh} and away side (or back-to-back) $\chi\approx\pi$~\cite{Moult:2018jzp,Duhr:2022yyp}. In both cases, perturbative calculations have been performed to high orders~\cite{Belitsky:2013ofa,Moult:2018jzp,Dixon:2018qgp,Luo:2019nig,Dixon:2019uzg,Henn:2019gkr,Duhr:2022yyp} and insights have also been gained from connections to conformal field theories~\cite{Hofman:2008ar,Belitsky:2013xxa,Belitsky:2013bja,Belitsky:2013ofa,Dixon:2019uzg,Kologlu:2019mfz,Chang:2020qpj,Chen:2020adz,Chen:2023zzh}.  However, only recently has there been a more intense focus on the nonperturbative region of EECs~\cite{Schindler:2023cww,Lee:2024esz,Chen:2024nyc,Kang:2024dja,Barata:2024wsu,Liu:2024lxy,Csaki:2024zig,Csaki:2024joe,Lee:2024jnt}. 

In this Letter, we establish how the free hadron and transition regions of the near-side EEC can be described using the language of dihadron fragmentation functions~(DiFFs).
We explicitly show how at large relative transverse momentum between the two hadrons, a nonperturbative function we call the ``EEC-DiFF'' reproduces the results of the ``EEC jet'' function in the quark/gluon region.  Therefore, this theoretical framework will serve as a bridge to allow the free hadron, transition, and quark/gluon regions to be analyzed simultaneously.  In addition, we develop a simple model for our EEC-DiFF based on the properties of the underlying DiFF, which allows  one to understand salient features of near-side EECs in a natural way. We then perform the first fit within the dihadron framework of $e^+e^-$ experimental measurements in the free hadron and transition regions, finding a reasonable description of the data. 

{\it EEC Dihadron Fragmentation Function---} Our starting point is the DiFF $D_1^{h_1h_2/q}(\xi_1,\xi_2,\vec{R}_T)$ defined in Ref.~\cite{Pitonyak:2023gjx} (see also Ref.~\cite{Pitonyak:2025lin}):
\begin{align}
    &\!\!D_1^{h_1h_2/q}(\xi_1,\xi_2,\vec{R}_T) \!=\! \frac{\xi^2}{64\pi^3\xi_1\xi_2}\!\int\! \!d^2\vec{k}_T\!\!\sum_X\hspace{-0.5cm}\int\! \int\!\!\frac{dx^+\!d^2\vec{x}_\perp}{(2\pi)^3}e^{ik\cdot x}\nonumber\\
    &\;\times {\rm Tr}\langle 0|\gamma^-\psi(x)|P_1,P_2;X\rangle\langle P_1,P_2;X|\bar{\psi}(0)|0\rangle\big|_{x^-=0}\,,
    \label{e:D1z1z2RT}
\end{align}
where $\xi_1,\xi_2$ are the fractions carried by the hadrons $h_1,h_2$ of the lightcone momentum of the fragmenting quark $q$ (with momentum $k$), and $\vec{R}_T\equiv \frac{1}{2}(\vec{P}_{1T}-\vec{P}_{2T})$ is (half of) their relative transverse momentum. (We have suppressed gauge links for brevity and a color average is understood.)  An analogous DiFF can be written down for gluon fragmentation~\cite{Pitonyak:2023gjx}.  The function~\eqref{e:D1z1z2RT} is a density in $(\xi_1,\xi_2,\vec{R}_T)$ for the number of hadron pairs $(h_1h_2)$ produced when a quark $q$ fragments~\cite{Pitonyak:2023gjx}.  The total momentum of the dihadron is denoted by $P_h$, with $M_h^2=P_h^2$ its invariant mass, and $\xi\equiv \xi_1+\xi_2$.  We work in the ``dihadron frame'' where $P_h$ has no transverse momentum and has a large lightcone-minus component.    

An important aspect of $D_1^{h_1h_2/i}(\xi_1,\xi_2,\vec{R}_T)$ ($i$ is a generic parton) is that it has been integrated over the transverse momentum $\vec{k}_T$ of the fragmenting parton.  This allows for {\it collinear} factorization theorems and evolution equations to be used whose partonic cross sections and splitting functions are the same as for single-hadron~FFs~\cite{Pitonyak:2025lin,Rogers:2024nhb}.  Such results do not hold for the fully unintegrated function that depends on both $\vec{k}_T$ and $\vec{R}_T$ (or $\vec{P}_{1\perp}$ and $\vec{P}_{2\perp}$ if one works in a frame where the parton has no transverse momentum). 

We also introduce the variable $z_{12}\equiv \frac{1}{2}\!\left(\!1-\tfrac{\vec{P}_1\cdot\vec{P}_2}{|\vec{P}_1||\vec{P}_2|}\right) = \frac{1}{2}(1-\cos\theta_{12})$, where $\theta_{12}$ is the angle between $h_1$ and $h_2$. Up to power suppressed (p.s.)~corrections 
$\sim 1/Q^2$, 
one can show for $e^+e^-\to h_1 h_2\,X$, 
\begin{equation}
    z_{12} = \frac{R_T^2}{Q^2}\frac{\tau^2}{\tau_1^2\tau_2^2}\,,
\end{equation}
with $R_T^2 \equiv \vec{R}_T^2$, $\tau_{1(2)}=2P_{1(2)}\!\cdot \!q/Q^2=2E_{1(2)}/Q$ and $\tau\equiv\tau_1+\tau_2$, where $E_{1(2)}$ is the energy of hadron $h_{1(2)}$, and $Q=\sqrt{q^2}$ is the center-of-mass (c.m.)~energy of the collision. 

We then define a nonperturbative function that we call the ``EEC-DiFF'' as
\begin{align}
\mathcal{D}^i(z_\chi, Q^2)\equiv\sum_{h_1,h_2}&\int \!\!d\xi_1 \!\int \!\!d\xi_2\int \!\!d^2\!\vec{R}_T\,\delta\!\left(\!z_\chi-\frac{R_T^2}{Q^2}\frac{\xi^2}{\xi_1^2\xi_2^2}\right)\nonumber\\
    &\times\, \xi_1\xi_2\,D_1^{h_1h_2/i}(\xi_1,\xi_2,\vec{R}_T)\,,\label{e:EEC_DiFF}
\end{align}
where $z_\chi\equiv \frac{1}{2}(1-\cos\chi)$.  (We are using $z_\chi$ instead of the standard $z$ as to not create confusion with the momentum fraction $z$ commonly used with FFs.) We assume that $\mathcal{D}^q$, while different than $\mathcal{D}^g$ for gluons, is flavor independent.  We focus on the terms at finite $z_\chi$ 
and omit contributions proportional to $\delta(z_\chi)$, which arise when both EEC detectors register the same particle~\cite{Dixon:2019uzg}. As we will show, the EEC-DiFF enters naturally into our theoretical calculation of ${\rm EEC}(\chi)$ valid for the near-side free hadron and transition regions and plays a crucial phenomenological role in explaining their behaviors.

In order to study experiments across various  c.m.~energies, an evolution equation for $\mathcal{D}^i(z_\chi,Q^2)$ is also needed.  This can be obtained from the evolution equation for $D_1^{h_1h_2/i}(\xi_1,\xi_2,\vec{R}_T)$, which is known to $\mathcal{O}(\alpha_s)$~\cite{Ceccopieri:2007ip,Pitonyak:2023gjx}.   
The splitting functions for the ``homogeneous'' terms in the evolution of $D_1^{h_1h_2/i}(\xi_1,\xi_2,\vec{R}_T)$ that involve the DiFFs themselves are the same as for the single-hadron FFs $D_1^{h/i}(\xi)$~\cite{Pitonyak:2025lin,Rogers:2024nhb} and are known to $\mathcal{O}(\alpha_s^3)$~\cite{Mitov:2006wy,Mitov:2006ic,Moch:2007tx,Almasy:2011eq}.  However, starting at $\mathcal{O}(\alpha_s^2)$, the evolution of $D_1^{h_1h_2/i}(\xi_1,\xi_2,\vec{R}_T)$ is expected to contain
``inhomogeneous'' terms that involve single-hadron FFs~\cite{deFlorian:2003cg,Majumder:2004br,Majumder:2004wh,Chen:2022pdu,Chen:2022muj,Pitonyak:2023gjx}, which have not been calculated yet. Therefore, we will approximate the evolution of $\mathcal{D}^i(z_\chi,Q^2)$ by neglecting this mixing with single-hadron FFs and focus only on the  homogeneous part involving DiFFs. As we derive in the Supplemental Material, the evolution equation reads
\begin{equation}
    \frac{\partial\vec{\mathcal{D}}(z_\chi,Q^2;\mu)}{\partial\ln\mu^2}\overset{{\rm DiFF}}{=}\int_0^1\!dw\,w^2\,\vec{\mathcal{D}}(z_\chi,w^2Q^2;\mu)\cdot\hat{P}_T(w;\mu)\,,\label{e:EECDiFF_evo}
\end{equation}
where $\vec{\mathcal{D}}\equiv (\mathcal{D}^q,\mathcal{D}^g)$ is a row vector that contains the quark and gluon EEC-DiFFs.    

{\it Connection to the EEC Quark/Gluon Region---} When $R_T\gtrsim \Lambda_{\rm QCD}$, one can assume the two hadrons emerged from a single parton fragmentation encoded by a DiFF like $D_1^{h_1h_2/i}(\xi_1,\xi_2,\vec{R}_T)$~\cite{Collins:1993kq,Collins:1994ax,Bianconi:1999cd,Rogers:2024nhb,Pitonyak:2025lin}.  Once $R_T\gg \Lambda_{\rm QCD}$, though, one must also consider contributions where the two hadrons were produced from two different parton fragmentations.  In this region, the DiFF $D_1^{h_1h_2/i}(\xi_1,\xi_2,\vec{R}_T)$ can be expanded for large $R_T$ into single-hadron FFs~\cite{Zhou:2011ba,Pitonyak:2023gjx}. 
A straightforward calculation gives 
\begin{align}
    &D_1^{h_1h_2/q}(\xi_1,\xi_2,\vec{R}_T) \overset{R_T\gg\Lambda_{\rm QCD}}{=} \frac{1}{R_T^2}\frac{C_F\alpha_s}{2\pi^2}\nonumber\\
    &\times\int_{\xi_1}^{1-\xi_2}\!\!\!\frac{dw}{w(1-w)}\!\!\left[\frac{1+w^2}{1-w}D_1^{h_1/q}(\tfrac{\xi_1}{w})D_1^{h_2/g}(\tfrac{\xi_2}{1-w})\right],
\end{align}
where we also implicitly include a term with $(h_1\leftrightarrow h_2)$ and omit a term proportional to $\delta^2(\vec{R}_T)$, in accordance with our earlier treatment, as it contributes only to $\delta(z_\chi)$. Using Eq.~\eqref{e:EEC_DiFF} along with the momentum sum rule $\sum_h \int \!dz\,z\,D_1^{h/i}(z)=1$, we find
\begin{equation}
    \mathcal{D}^q(z_\chi,Q^2)\overset{R_T\gg\Lambda_{\rm QCD}}{=}\frac{\alpha_s}{4\pi}\frac{3C_F}{z_\chi}\,.
\end{equation}
This agrees exactly with the $\mathcal{O}(\alpha_s)$ term for the ``EEC jet'' function $dJ_q(z_\chi)/dz_\chi$ in Ref.~\cite{Dixon:2019uzg} that underlies ${\rm EEC}(\chi)$ in the quark/gluon (perturbative) region.  We verified that the same occurs with the gluon EEC-DiFF $\mathcal{D}^g(z_\chi)$ and gluon EEC jet function $dJ_g(z_\chi)/dz_\chi$.  Thus, within the DiFF framework, an analytical matching will be able to bridge the near-side free hadron, transition, and quark/gluon regions and allow all of them to be analyzed simultaneously. The connection between the EEC-DiFF and the EEC jet functions also provides theoretical support for the universality of EEC jet functions and the analyses presented in Refs.~\cite{Barata:2024wsu,Herrmann:2025fqy}.

{\it Theoretical Result for EECs in $e^+e^-$ Annihilation---} The energy-energy correlator in $e^+e^-\to h_1 h_2\,X$ for two hadrons separated by an angle $\chi$ is defined as 
\begin{equation}
    {\rm EEC(\chi)}\equiv \frac{1}{\sigma_t}\frac{d\Sigma}{d\chi}=\frac{\sin\chi}{2} \frac{1}{\sigma_t}\!\sum_{h_1,h_2}\int \!\!d\sigma\frac{E_1E_2}{Q^2}\,\delta(z_\chi-z_{12})\,,\label{e:EEC_cs}
\end{equation}
where $d\sigma$ is the differential cross section for $e^+e^-\to h_1 h_2\,X$, and $\sigma_t$ is the total cross section for $e^+e^-\to {\rm hadrons}$. As we derive in the Supplemental Material, in a region where the cross section is only due to dihadron fragmentation, ${\rm EEC}(\chi)$ is given by 
\begin{align} \label{e:EEC}
    &{\rm EEC}(\chi)\overset{{\rm DiFF}}{=}\frac{\sin\chi}{2}\frac{\sigma_0}{\sigma_t}\!
    \int_0^1 \!\!dw\,w^2 \,\vec{\mathcal{D}}(z_\chi, w^2Q^2;\mu)\!\cdot\!\vec{H}(w;\tfrac{\mu}{Q})\,,
\end{align}
where $\vec{H}\equiv (H^{q},H^g)$ are hard factors for $e^+e^-\to h\,X$, which are known to $\mathcal{O}(\alpha_s^3)$~\cite{Mitov:2006ic,Moch:2007tx,Almasy:2011eq,He:2025hin}, and $\sigma_0$ is the  Born-level cross section for $e^+e^-\to {\rm hadrons}$.  

{\it Model for the EEC-DiFF---}  In order to analyze experimental measurements on near-side EECs in $e^+e^-$ annihilation, we need a model for the EEC-DiFF $\mathcal{D}^i(z_\chi,Q^2)$. To do so, we utilize properties of the DiFF $D_1^{h_1h_2/i}(\xi_1,\xi_2,\vec{R}_T)$ that enters Eq.~\eqref{e:EEC_DiFF} and underlies $\mathcal{D}^i(z_\chi,Q^2)$.  Evaluating the $R_T$-integration in Eq.~\eqref{e:EEC_DiFF} with the delta function generates a prefactor proportional to $Q^2$ and fixes $R_T^2\!\!=\!\!z_\chi Q^2 \!\left(\tfrac{\xi_1^2\xi_2^2}{\xi^2}\right)$.  We also assume $D_1^{h_1h_2/i}(\xi_1,\xi_2,\vec{R}_T)$ to have a Gaussian dependence on $R_T\sim \sqrt{z_\chi}\, Q$, analogous to the simple ansatz used in a parton model framework for the transverse momentum dependence of TMDs~\cite{Anselmino:2005nn,Signori:2013mda,Anselmino:2013lza}. Thus, the observed increase in the peak value of the near-side EEC with $Q$, its location at roughly the same $\sqrt{z_\chi}\,Q\approx \chi Q/2$ for all $Q$, as well as the scaling of ${\rm EEC}(\chi)$ with $Q^2$ as $\chi\to 0$~\cite{CMS:2024mlf,ALICE:2024dfl,Liu:2024lxy,Barata:2024wsu,Chen:2025rjc} can all naturally be understood on theoretical grounds alone within the DiFF framework.  

Our model for $\vec{\mathcal{D}}(z_\chi,Q^2;\mu_0)$ at the initial scale $\mu_0=m_b=4.18\,{\rm GeV}$ then reads 
\begin{equation}
    \mathcal{D}^q(z_\chi,Q^2;\mu_0) = NQ^2\,\frac{\exp(-z_\chi Q^2/a)}{1+z_\chi Q^2/b}\,,
\end{equation}
where $\{N,a,b\}$ are free parameters to be fit to the data. The denominator introduces a deviation from pure Gaussian behavior at larger values of $\sqrt{z_\chi}\,Q$, allowing for a more accurate description of the transition region.
We set $\mathcal{D}^g(z_\chi,Q^2;\mu_0)=0$, allowing it to become nonzero through evolution. Since the EEC-DiFF is assumed to be flavor independent, we have chosen $\mu_0=m_b$ so that the evolution starts with all five flavors active.

{\it Data Selection---} 
An important consideration in our study is what cut to make on the data to ensure we are in a region dominated by DiFFs.  In this regard, a recent extraction of DiFFs~\cite{Cocuzza:2023vqs} from a Belle cross section measurement~\cite{Belle:2017rwm} and PYTHIA data can serve as guidance.  To be specific, Ref.~\cite{Cocuzza:2023vqs} was able to successfully describe data from $Q=10.6\,{\rm GeV}$ to $91.2\,{\rm GeV}$ on $\pi^+\pi^-$ final states within a leading-order DiFF framework that involved the function $D_1^{\pi^+\pi^-/i}(\tau,M_h)$.  For all c.m.~energies, the data satisfied $M_h/\tau \lesssim 3.5\,{\rm GeV}$.  Since $R_T=\frac{\sqrt{\tau_1\tau_2}}{\tau}M_h$ up to p.s.~corrections, and recall $\sqrt{z_\chi}=\frac{R_T}{Q}\frac{\tau}{\tau_1\tau_2}$, one then has the condition $\sqrt{z_\chi}\,Q\sqrt{\tau_1\tau_2}/\tau\lesssim 3.5\,{\rm GeV}$. Assuming $\tau_1 \approx \tau_2 \approx \tau/2$ for most near-side EEC events, we arrive at a practical data cut of $\sqrt{z_\chi}\,Q < 7\,\mathrm{GeV}$. For the range of $Q$ values in Table~\ref{t:chi2}, this condition corresponds to angular separations approximately within $1.5^\circ \lesssim \chi \lesssim 34^\circ$. Therefore, one should not interpret ``large angles'' as necessarily corresponding to the perturbative (quark/gluon) region. Rather, as also emphasized in a factorization-based analysis of EECs~\cite{Dixon:2019uzg}, it is the size of the scale $\sqrt{z_\chi}\,Q$ (or, equivalently, $R_T$ or $M_h$) that determines the physical regime being probed.

In fitting our model parameters, we used datasets with $Q \ge 22\,{\rm GeV}$ that included collectively both charged and neutral particles, provided statistical and systematic uncertainties or their quadrature sum, and had more than one bin after the cut $\sqrt{z_\chi}\,Q < 7\,{\rm GeV}$.  This left us with the following datasets for our study:~TASSO~\cite{TASSO:1987mcs}, MAC~\cite{Fernandez:1984db}, MARKII~\cite{Wood:1987uf}, TOPAZ~\cite{TOPAZ:1989yod}, OPAL~\cite{OPAL:1991uui}.   For measurements whose uncertainties were rounded to one significant digit, we expand them to two significant digits assuming maximal uncertainty before rounding, following the procedure in Ref.~\cite{Kardos:2018kqj}.

{\it Phenomenological Methodology and Results for EECs in $e^+e^-$ Annihilation---} 
We work at leading order in the hard factors, $H^{q}(w)=\tfrac{1}{2}\delta(1-w),H^g(w)=0$ in Eq.~\eqref{e:EEC}, and leading-log evolution, where $w$ is set to 1 in the argument of $\vec{\mathcal{D}}$ on the r.h.s.~of Eq.~\eqref{e:EECDiFF_evo}.  We use a Bayesian Monte Carlo method with data resampling to obtain the posterior distributions of the parameters $\{N,a,b\}$.  We find a better description of the data is achieved evolving $\vec{\mathcal{D}}(z_\chi,Q^2;\mu)$ to $\mu=0.5Q$ instead of $\mu=Q$, which can be associated with the fact that the dihadron will only carry some fraction of the energy of the parton that initiated the jet~\cite{Liu:2024lxy}.   
We also account for the uncertainty in the choice of scale by varying it to $\mu=0.5Q/2$ and $\mu=2\cdot 0.5Q$.  For each scale setting ($\mu=0.5Q,\,0.5Q/2,\,2\cdot 0.5Q$) we gather 150 replicas.  From the collective 450 replicas we compute the mean and standard deviation and use them to generate a central curve and 1$\sigma$ error band for our theory calculation.

\begin{figure}[b!]
\includegraphics[width=0.48\textwidth]{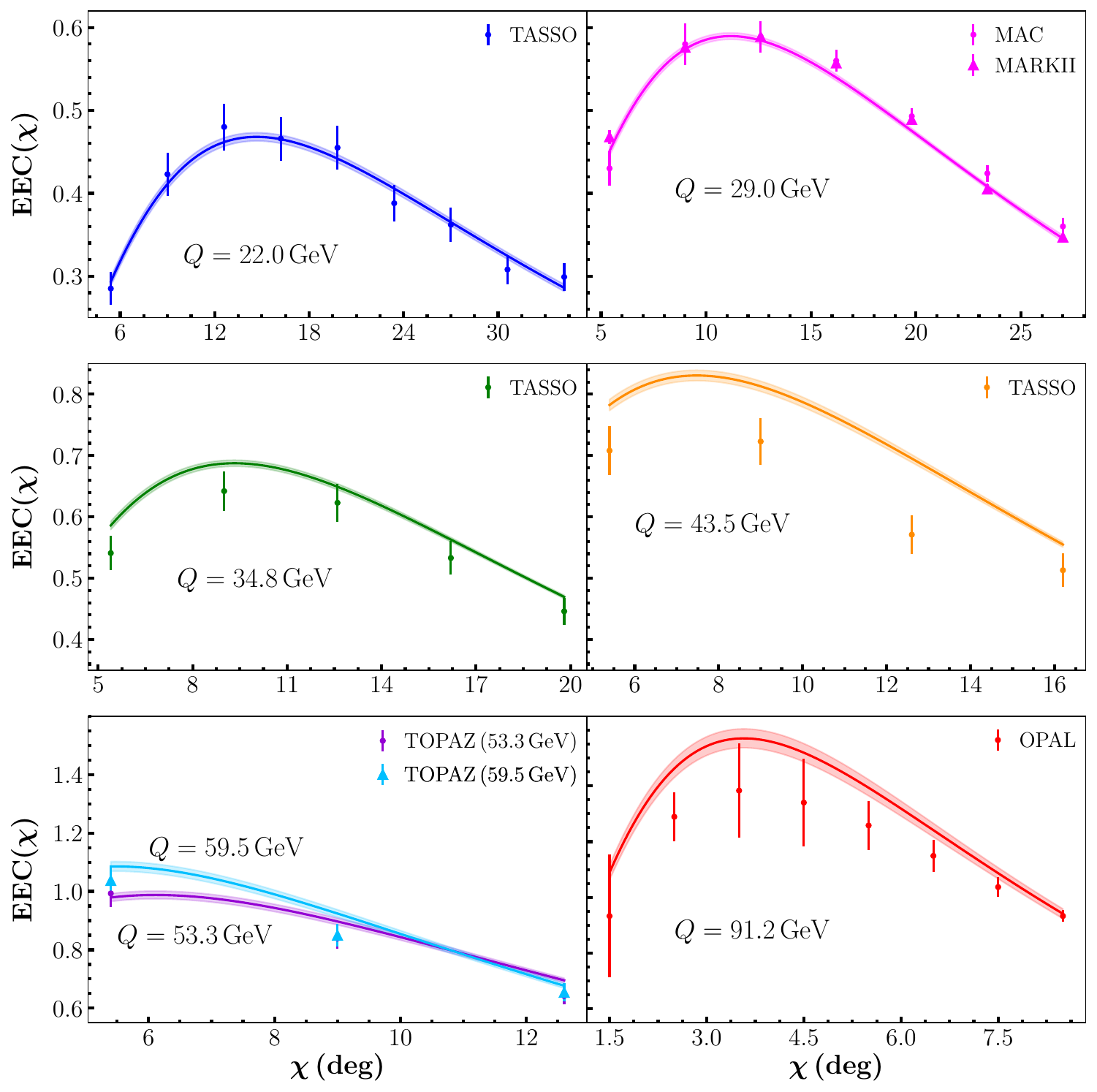}
\vspace{-0.5cm}
\caption
{${\rm EEC}(\chi)$ vs.~$\chi$ for our theoretical central curve and $1\sigma$ error band compared to the data used in our analysis.}
\label{f:theory}
\vspace{-0.35cm}
\end{figure}

The results for ${\rm EEC}(\chi)$ vs.~$\chi$ are shown in Fig.~\ref{f:theory}, where we find reasonable agreement with the data, with $\chi^2/N_{\rm pts}=86.4/46 = 1.88$ and $Z$-score $=3.44$, where $Z\equiv \sqrt{2}~{\rm erf}^{-1}(2p-1)$ with the $p$-value computed using the total $\chi^2$.  Table~\ref{t:chi2} summarizes the $\chi^2/N_{\rm pts}$ for each dataset. We see that the somewhat larger $\chi^2/N_{\rm pts}$ is mostly due to the TASSO $43.5\,{\rm GeV}$ data.  Given that the other TASSO datasets at $22\,{\rm GeV}$ and $34.8\,{\rm GeV}$, as well as measurements by other collaborations at higher energies, are well described, there could be a normalization issue with the TASSO $43.5\,{\rm GeV}$ measurement.  Higher-order corrections to the evolution of the EEC-DiFF and hard factors in Eq.~(8) could improve the $\chi^2/{N_{\rm pts}}$~\cite{Herrmann:2025fqy}.

\begin{table}[t!]
    \begin{tabular}{c c c c} 
    \hline
    {\bf Collaboration }& 
        ~$\boldsymbol{Q\,{\rm (GeV)}}$~ & 
        ~$\boldsymbol{N_{\rm pts}}$ &
        ~$\boldsymbol{\chi^2 / N_\mathrm{pts}}$~  
        \\ \hline\hline
    TASSO   \cite{TASSO:1987mcs} &
        $22.0$ &
        $9$ &
        $0.35$
        \\ \hline
    MAC   \cite{Fernandez:1984db} &
        $29.0$ &
        $7$ &
        $1.69$
        \\ \hline
    MARKII   \cite{Wood:1987uf} &
        $29.0$ &
        $7$ &
        $1.85$
        \\ \hline
    TASSO   \cite{TASSO:1987mcs} &
        $34.8$ &
        $5$ &
        $1.42$
        \\ \hline
    TASSO   \cite{TASSO:1987mcs} &
        $43.5$ &
        $4$ &
        $6.71$
        \\ \hline
    TOPAZ   \cite{TOPAZ:1989yod} &
        $53.3$ &
        $3$ &
        $1.91$
        \\ \hline
    TOPAZ   \cite{TOPAZ:1989yod} &
        $59.5$ &
        $3$ &
        $1.71$
        \\ \hline
    OPAL   \cite{OPAL:1991uui} &
        $91.2$ &
        $8$ &
        $1.70$
        \\ \hline\hline
    {\bf Total} &  & 46 & 1.88 \\ 
    & & & ($Z=3.44$)\\ \hline
    \end{tabular}
\caption{Summary of the $\chi^2/N_{\rm pts}$ for each dataset.}
\label{t:chi2}
\end{table}

\begin{figure}[t!]
\includegraphics[width=0.48\textwidth]{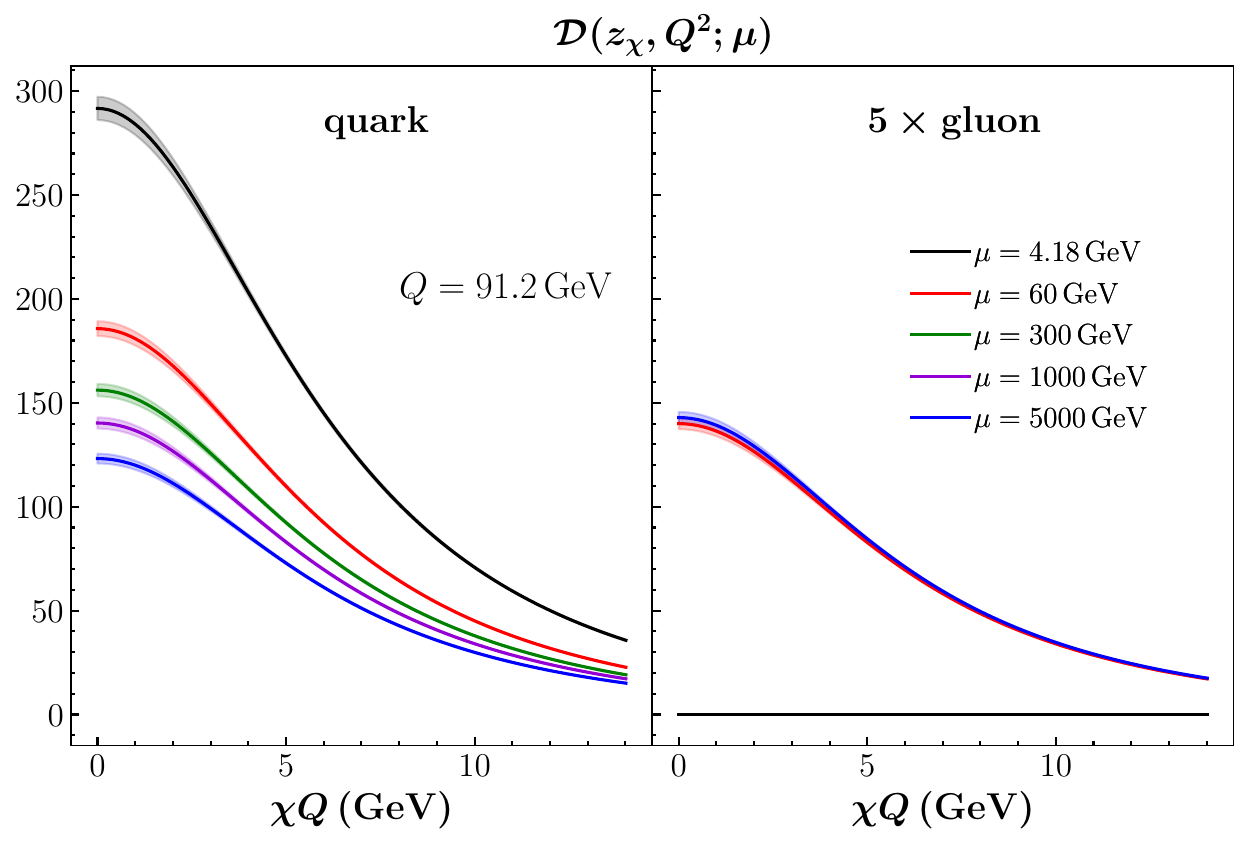}
\vspace{-0.5cm}
\caption
{$\mathcal{D}^i(z_\chi,Q^2;\mu)$ vs.~$\chi Q$ with a $1\sigma$ error band for the quark~(left) and gluon (right) at fixed $Q=91.2\,{\rm GeV}$ for various values of $\mu$. Note that the gluon curves have been multiplied by a factor of five and the inset zooms in on $0<\chi Q<2\,{\rm GeV}$.}
\label{f:EEC_DiFF}
\vspace{-0.35cm}
\end{figure}

The extracted EEC-DiFF $\mathcal{D}^i(z_\chi,Q^2;\mu)$ is displayed in Fig.~\ref{f:EEC_DiFF} at fixed $Q=91.2\,{\rm GeV}$  for various values of~$\mu$. As expected, the quark contribution is significantly larger than the gluon, which was initialized to zero at the starting scale. The fitted parameter values for $\mathcal{D}^q(z_\chi,Q^2;\mu_0)$ are $N=0.035\pm0.003\,{\rm GeV^{-2}}, a=150\pm 19\,{\rm GeV^{2}}, b=10.1\pm 0.6\,{\rm GeV^{2}}$. Since $\mathcal{D}^i(z_\chi, Q^2; \mu)$ is constructed from the underlying DiFF $D_1^{h_1 h_2/i}(\tau_1, \tau_2, \vec{R}_T)$ via Eq.~\eqref{e:EEC_DiFF}, it may be possible to independently constrain or validate the EEC-DiFF through a dedicated $e^+e^-$ cross section measurement of $d\sigma/d\tau_1 d\tau_2 dR_T^2$.

\begin{figure}[h!]
\includegraphics[width=0.48\textwidth]{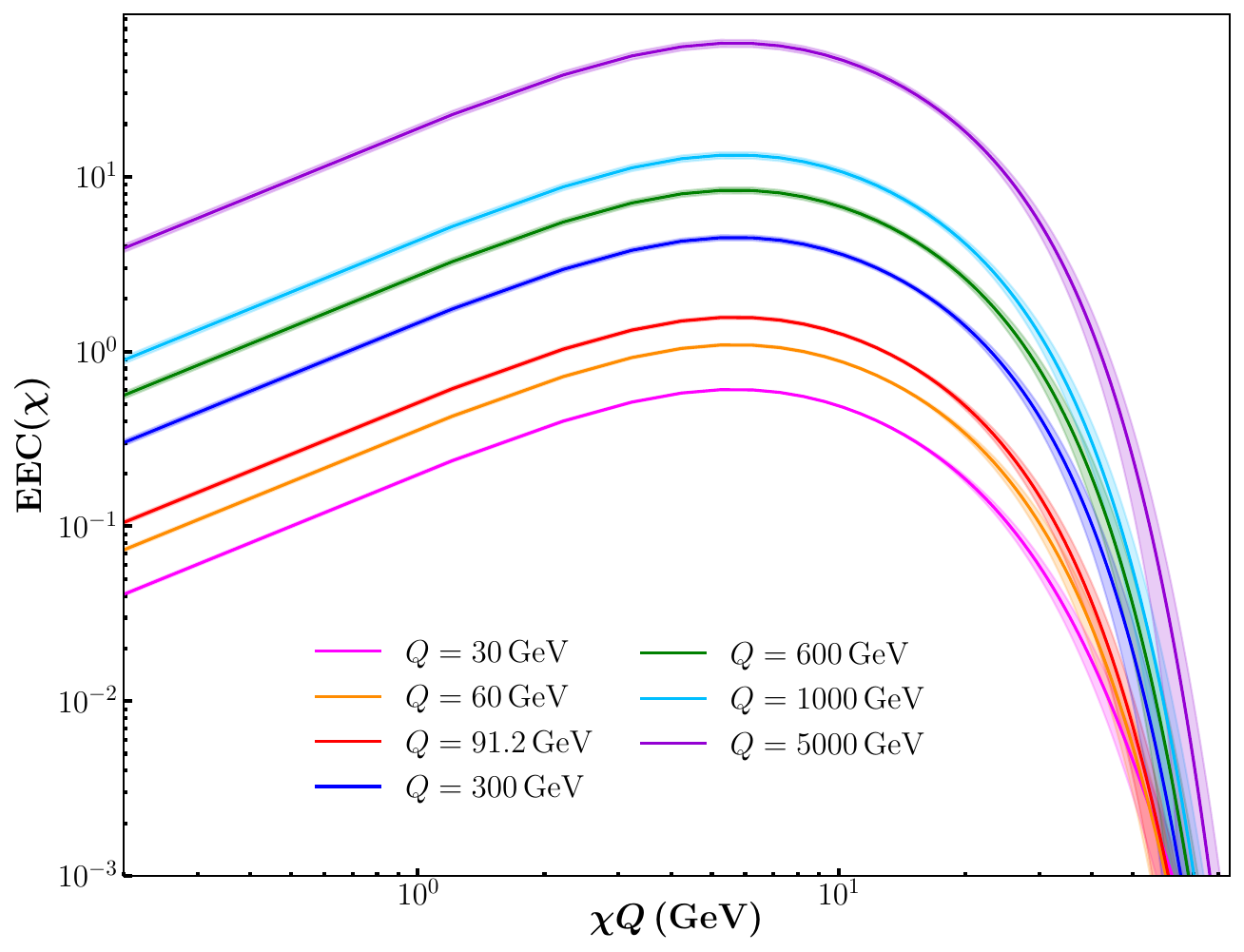}
\vspace{-0.5cm}
\caption
{${\rm EEC}(\chi)$ vs.~$\chi Q$ with a $1\sigma$ error band for various values of $Q$.}
\label{f:EEC_chiQ}
\vspace{-0.35cm}
\end{figure}

In Fig.~\ref{f:EEC_chiQ}, we plot ${\rm EEC}(\chi)$ as a function of $\chi Q$ to explicitly demonstrate that our phenomenological analysis reproduces the key features of the near-side EEC anticipated from DiFF-based theoretical arguments. Specifically, we observe that the peak value of the near-side EEC increases with $Q$, while its position remains approximately fixed in $\chi Q$ across all $Q$ values. We also confirmed that the curves for ${\rm EEC}(\chi)/Q^2$ versus $\chi$ tend to converge as $\chi \to 0$, consistent with the expected scaling behavior~\cite{Chen:2025rjc}. 
Overall, our EEC-DiFF framework quantitatively captures the main features of the near-side free hadron and transition regions, while simultaneously providing a formal theoretical foundation that enables matching to the quark/gluon regime.

{\it Summary and Outlook---} We have established a framework, using dihadron fragmentation, for analyzing the free hadron and transition regions of near-side EECs. We explicitly showed to $\mathcal{O}(\alpha_s)$ that at large relative transverse momentum of the two hadrons, an agreement occurs between the EEC-DiFF in Eq.~\eqref{e:EEC_DiFF} and the EEC jet function used in the quark/gluon region~\cite{Dixon:2019uzg}. Therefore, it will be possible in the future to match the free hadron, transition, and quark/gluon regions in this theoretical framework, allowing all of them to be analyzed simultaneously.  Using a simple model for the EEC-DiFF, we have been able to successfully describe for the first time within the dihadron framework near-side EEC measurements in the free hadron and transition regions across a range of center-of-mass energies. The DiFF approach also naturally explains the $Q$ dependence of ${\rm EEC}(\chi)$ in this regime and its scaling behaviors. The DiFF formalism for EECs can provide new insights into hadronization in QCD and, due to its close connection with transverse spin physics~~\cite{Bianconi:1999uc,Radici:2001na,Bacchetta:2002ux,Bacchetta:2003vn,Boer:2003ya,Bacchetta:2004it,Bacchetta:2008wb,Bacchetta:2011ip,Courtoy:2012ry,Courtoy:2012ry,Bacchetta:2012ty,Matevosyan:2013aka,Gliske:2014wba,Radici:2015mwa, Radici:2016lam,Matevosyan:2017alv,Matevosyan:2017uls,Matevosyan:2017liq,Radici:2018iag,Matevosyan:2018jht,Matevosyan:2018icf,Benel:2019mcq,Cocuzza:2023oam,Cocuzza:2023vqs}, will also open up an avenue to explore azimuthal/spin-dependent near-side EEC observables in $e^+e^-$, lepton-nucleon, and proton-proton collisions.

{\it Note Added}:~While finalizing this manuscript, Refs.~\cite{Lee:2025okn,Guo:2025zwb,Chang:2025kgq} appeared as preprints. Our main results and conclusions are consistent with theirs, and, in addition, we have performed a phenomenological fit to existing $e^+e^-$ experimental data.

Z.K. thanks the DESY theory group for their hospitality during the course of this work. This work was supported by the National Science Foundation under Grants  No.~PHY-2515057 (Z.K., C.Z.), No.~PHY-2110472 and No.~PHY-2412792 (A.M.), and No.~PHY-2308567 (D.P.). 

{\it Data Availability}:~The data that support the findings of this article are openly available~\cite{EECDiFF_git}.

\bibliography{dp}

\clearpage
\onecolumngrid
\section{Supplemental Material}
\setcounter{equation}{0}
\renewcommand{\thesubsection}{S\arabic{subsection}}   
\setcounter{secnumdepth}{2}
\renewcommand{\thetable}{S\arabic{table}}   
\renewcommand{\thefigure}{S\arabic{figure}}
\renewcommand{\theequation}{S\arabic{equation}}

\subsection*{Dihadron Fragmentation Result for EEC($\chi$) in $e^+e^-$ Annihilation and Evolution Equation for the EEC-DiFF}

From Refs.~\cite{Rogers:2024nhb, Pitonyak:2025lin} we know that the terms in the cross section $(d\sigma/d\tau_1 d\tau_2 d^2\!\vec{R}_T)^{e^+e^-\to h_1h_2X}$ that only involve dihadron fragmentation will have exactly the same structure as $(d\sigma/d\tau)^{e^+e^-\to hX}$ with $D_1^{h/i}\to D_1^{h_1h_2/i}$. 
However, the replacement $dw/w\to dw/w^2$ is also needed -- see, e.g., Eq.~(16b) of Ref.~\cite{Pitonyak:2025lin}.  Notably, the partonic cross sections are the same in both cases.  Based on the known factorization formula for $(d\sigma/d\tau)^{e^+e^-\to hX}$~\cite{Mitov:2006ic,Moch:2007tx,Almasy:2011eq,He:2025hin}, one can then immediately write down 
\begin{align}
    \frac{d\sigma}{d\tau_1 d\tau_2 d^2\!\vec{R}_T} \overset{{\rm DiFF}}{=} \sum_{k=T,L}\int_\tau^1\! \frac{dw}{w^2}\bigg\{&\sigma_0\bigg[D_{1,{\rm S}}^{h_1h_2}(\tfrac{\tau_1}{w},\tfrac{\tau_2}{w},\vec{R}_T;\mu)\,\mathbb{C}_{k,q}^{\rm S}(w;\mu/Q)+D_1^{h_1h_2/g}(\tfrac{\tau_1}{w},\tfrac{\tau_2}{w},\vec{R}_T;\mu)\,\mathbb{C}_{k,g}^{\rm S}(w;\mu/Q)\bigg]\nonumber\\
    &+\sum_q\sigma_q^0\,D^{h_1h_2/q}_{1,{\rm NS}}(\tfrac{\tau_1}{w},\tfrac{\tau_2}{w},\vec{R}_T;\mu)\,\mathbb{C}_{k,q}^{\rm NS}(w;\mu/Q)\bigg\}\,,\label{e:DiFFcs}
\end{align}
where $\mathbb{C}_{k,i}^{\rm S(NS)}$ are the flavor singlet (non-singlet) coefficient functions for the transverse ($T$) and longitudinal ($L$) structure functions, $D_{1,{\rm S}}^{h_1h_2}\equiv\frac{1}{N_f}\sum_q(D_1^{h_1h_2/q}+D_1^{h_1h_2/\bar{q}})$, and $D_{1,{\rm NS}}^{h_1h_2/q}\equiv(D_1^{h_1h_2/q}+D_1^{h_1h_2/\bar{q}}-D_{1,{\rm S}}^{h_1h_2})$. The Born-level cross section for $e^+e^-\to {\rm hadrons}$ is $\sigma_0\equiv \sum_q\sigma_q^0$, with $\sigma_q^0\equiv (4\pi\alpha_{em}^2N_c\bar{e}_q^2)/(3Q^2)$, where $\bar{e}_q$ accounts for electroweak interactions. 

Substituting Eq.~\eqref{e:DiFFcs} into Eq.~\eqref{e:EEC_cs}, with $d\sigma= d\tau_1d\tau_2d^2\!\vec{R}_T\cdot (d\sigma/d\tau_1d\tau_2d^2\!\vec{R}_T)$, employing the definition of $\mathcal{D}^i$ in Eq.~\eqref{e:EEC_DiFF}, and assuming the flavor independence of $\mathcal{D}^q$, we arrive at
\begin{align}
    {\rm EEC}(\chi)&\overset{{\rm DiFF}}{=} \frac{\sin\chi}{2}\frac{\sigma_0}{\sigma_t}\sum_{k=T,L}\bigg[\frac{1}{2}\int_0^1 \!dw\,w^2\,\mathcal{D}^q(z_\chi, w^2Q^2;\mu)\,\mathbb{C}^{\rm S}_{k,q}(w;\tfrac{\mu}{Q})+\frac{1}{4}\int_0^1 \!dw\,w^2\,\mathcal{D}^g(z_\chi, w^2Q^2;\mu)\,\mathbb{C}^{\rm S}_{k,g}(w;\tfrac{\mu}{Q})\bigg]\nonumber\\[0.3cm]
    &\overset{{\rm DiFF}}{=}\frac{\sin\chi}{2}\frac{\sigma_0}{\sigma_t}
    \int_0^1 \!dw\,w^2\bigg[\mathcal{D}^q(z_\chi, w^2Q^2;\mu)\,H^q(w;\tfrac{\mu}{Q})+\mathcal{D}^g(z_\chi, w^2Q^2;\mu)\,H^g(w;\tfrac{\mu}{Q})\bigg],\label{e:EEC_SM}
\end{align}
where we have defined $H^q(w;\tfrac{\mu}{Q})\equiv\frac{1}{2}\sum_{k=T,L}\mathbb{C}^{\rm S}_{k,q}(w;\tfrac{\mu}{Q})$ and $H^g(w;\tfrac{\mu}{Q})\equiv \frac{1}{4}\sum_{k=T,L}\mathbb{C}^{\rm S}_{k,g}(w;\tfrac{\mu}{Q})$ -- see also Ref.~\cite{Dixon:2019uzg}.  This is Eq.~\eqref{e:EEC} in the main text.

The evolution of $\mathcal{D}^i(z_\chi,Q^2)$  follows from the evolution of $D_1^{h_1h_2/i}(\xi_1,\xi_2,\vec{R}_T)$.  Because we assume flavor independence for $\mathcal{D}^q$, we can use the flavor singlet evolution equation on the row vector $\vec{D}_1^{h_1h_2}\equiv (D_1^{h_1h_2/q},D_1^{h_1h_2/g})$.  Similar to the cross section, as long as we only consider terms involving DiFFs, the evolution equation for $\vec{D}_1^{h_1h_2}$ will be the same as for single-hadron FFs but with $dw/w\to dw/w^2$ (see Eq.~(15b) of Ref.~\cite{Pitonyak:2025lin}).  Specifically,
\begin{equation}
    \frac{\partial\vec{D}_1^{h_1h_2}(\xi_1,\xi_2,\vec{R}_T;\mu)}{\partial\ln\mu^2} \overset{\rm DiFF}{=} \int_\xi^1\frac{dw}{w^2}\,\vec{D}_1^{h_1h_2}(\tfrac{\xi_1}{w},\tfrac{\xi_2}{w},\vec{R}_T;\mu)\cdot \hat{P}_T(w;\mu)\,,\label{e:D1evo}
\end{equation}
where $\hat{P}_T$ is the singlet timelike splitting kernel matrix:~$\hat{P}_T\equiv\begin{pmatrix} P_{qq} & P_{qg} \\ P_{gq} & P_{gg}\end{pmatrix}$~\cite{Mitov:2006wy,Mitov:2006ic,Moch:2007tx,Almasy:2011eq}.
(The $\mu$ dependence in $\hat{P}_T(w;\mu)$ is from its dependence on $\alpha_s(\mu)$.)   
Applying $\partial/\partial\ln\mu^2$ to both sides of Eq.~\eqref{e:EEC_DiFF} and using Eq.~\eqref{e:D1evo} yields,
\begin{align}
    \frac{\partial\vec{\mathcal{D}}(z_\chi,Q^2;\mu)}{\partial\ln\mu^2}&\overset{{\rm DiFF}}{=}\int_0^1\!dw\,w^2\,\vec{\mathcal{D}}(z_\chi,w^2Q^2;\mu)\cdot\hat{P}_T(w;\mu) \,.\label{e:EECDiFF_evo_SM}
\end{align}
This is Eq.~\eqref{e:EECDiFF_evo} in the main text.

\end{document}